\documentclass[12pt]{article}

\usepackage{amssymb}
\usepackage{amsmath}

\usepackage{epsfig}

\begin{document} %

\title{\bf Randall-Sundrum scenario with a small curvature and Drell-Yan
process at the LHC}

\author{A.V. Kisselev\thanks{Electronic address:
alexandre.kisselev@ihep.ru} \\
{\small Institute for High Energy Physics, 142281 Protvino, Russia}
\\
{\small and}
\\
{\small Department of Physics, Moscow State University, 119991
Moscow, Russia}}

\date{}

\maketitle

\thispagestyle{empty}



\begin{abstract}
The Randall-Sundrum-like scenario with the small curvature $\kappa$
(RSSC model) is studied in detail in comparison with the original
RS1 model. In the framework of the RSSC model, the $p_{\perp}$
distributions for the dilepton production at the LHC are calculated.
Both dielectron and dimuon events are taken into account. The
important feature of calculations is the account of the widths of
massive graviton excitations. For the summary statistics taken at 7
TeV ($L = 5 \ \mathrm{fb}^{-1}$) and 8 TeV ($L = 20 \
\mathrm{fb}^{-1}$), the exclusion limit on the 5-dimensional gravity
scale $M_5$ is set to be 6.84 TeV at 95$\%$ C.L. For $\sqrt{s} = 13$
TeV and integrated luminosity 30 fb$^{-1}$, the LHC search limit is
found to be 10.16 TeV. These bounds on $M_5$ are independent of
$\kappa$ (up to powerlike corrections), provided $\kappa \ll M_5$.
\end{abstract}





\section{Randall-Sunrum scenario with the small \\ curvature (RSSC model)}
\label{sec:1}

In a recent paper \cite{Kisselev:13}, the $p_{\perp}$ distributions
for dimuon production at the LHC were calculated in the framework of
the Randall-Sundrum-like scenario with a \emph{small curvature}
(RSSC model, in what follows). The LHC discovery limits on
5-dimensional gravity scale $M_5$ were obtained for both 7 TeV and
14 TeV. In the present paper, the combined analysis of \emph{both
dielectron and dimuon} events at 7, 8, and 13 TeV will be done.

Before presenting results of calculations for the Drell-Yan (DY)
process at the LHC, it is useful to compare the RSSC model with the
standard Randall-Sundrum model with two 3D branes (RS1 model
\cite{Randall:99}). The classical action of this scenario is given
by%
\footnote{Contrary to Ref.~\cite{Randall:99} our constants $\Lambda$
and $\Lambda_{1,2}$ are redefined.}
\begin{align}\label{RS_action}
S &= 2 \bar{M}_5^3 \int \!\! d^4x \!\! \int_{-\pi r_c}^{\pi r_c}
\!\! dy \, \sqrt{|G|} \, (\mathcal{R}
- \Lambda) \nonumber \\
&+ \int \!\! d^4x \sqrt{|g^{(1)}|} \, (\mathcal{L}_1 - 2 \bar{M}_5^3
\Lambda_1) + \int \!\! d^4x \sqrt{|g^{(2)}|} \, (\mathcal{L}_2 - 2
\bar{M}_5^3 \Lambda_2) \;,
\end{align}
where $G_{MN}(x,y)$ is the 5-dimensional metric, with $M,N =
\{\mu,4\}$; $\mu = 0,1,2,3$; $y$ is the 5-th dimension coordinate;
and $r_c$ is the size of the ED. The quantities
\begin{equation}
g^{(1)}_{\mu\nu}(x) = G_{\mu\nu}(x, y=0) \;, \quad
g^{(2)}_{\mu\nu}(x) = G_{\mu\nu}(x, y=\pi r_c)
\end{equation}
are induced metrics on the branes, and $\mathcal{L}_1$ and
$\mathcal{L}_2$ are brane Lagrangians. It is assumed that the
cosmological constant $\Lambda$ is negative. Thus, we have a slice
of the AdS$_5$ space-time.

In Ref.~\cite{Randall:99} the background warped metric was found to
be
\begin{equation}\label{RS_metric}
\mathrm{RS1}: \quad ds^2 = e^{-2 \kappa |y|} \, \eta_{\mu \nu} \,
dx^{\mu} \, dx^{\nu} - dy^2 \;,
\end{equation}
where $\eta_{\mu\nu}$ is the Minkowski tensor with the signature
$(1,-1,-1,-1)$. The periodicity $y = y + 2\pi r_c$ is imposed and
the points $(x_{\mu}, y)$ and $(x_{\mu}, -y)$ are identified. So,
one gets the orbifold $S^1\!/Z_2$.

In the RS1 model the 3D branes are located at the fixed points $y =
0$ (Plank brane) and $y = \pi r_c$ (TeV brane). The SM fields are
constrained to the TeV brane, while the gravity propagates in all
spatial dimensions.

The hierarchy relation between the 5-dimensional \emph{reduced}
gravity scale $\bar{M}_5$ and \emph{reduced} Planck mass
$\bar{M}_{\mathrm{Pl}}$ looks like \cite{Randall:99},
\begin{equation}\label{RS_hierarchy_relation}
\mathrm{RS1}: \quad \bar{M}_{\mathrm{Pl}}^2 =
\frac{\bar{M}_5^3}{\kappa} \left(1 - e^{-2 \pi \kappa r_c} \right)
\;.
\end{equation}
The reduced scales in \eqref{RS_hierarchy_relation} are defined as
follows
\begin{equation}\label{Pl_scale_reduced}
\bar{M}_{\mathrm{Pl}} = M_{\mathrm{Pl}} /\sqrt{8\pi} \simeq 0.20 \,
M_{\mathrm{Pl}} \simeq 2.4\cdot 10^{18} \ \mathrm{GeV} \;,
\end{equation}
\begin{equation}\label{grav_scale_reduced}
\bar{M}_5 = M_5 /(2\pi)^{1/3} \simeq 0.54 \, \bar{M}_5 \;.
\end{equation}

The warp factor $F = \exp(-2\kappa|y|)$ has the following values on
the branes
\begin{equation}\label{warp_factors_RS}
\mathrm{RS1}: \quad F\Big|_{y = 0} = 1 \;, \quad F\Big|_{y = \pi
r_c} = e^{- 2\pi \kappa r_c } \;.
\end{equation}
For $\kappa > 0$, we get for boundary cosmological terms
\begin{equation}\label{tensions_RS}
\mathrm{RS1}: \quad \Lambda_1 > 0 \;, \quad \Lambda_2 < 0 \;.
\end{equation}
Thus, the Planck brane has a positive tension, while the TeV brane
has a negative tension.

In order for the hierarchy relation \eqref{RS_hierarchy_relation} to
be satisfied, one has to put
\begin{equation}\label{RS_parameters}
\mathrm{RS1}: \quad \kappa \sim \bar{M}_5 \sim M_{\mathrm{Pl}} \;.
\end{equation}
The RS1 model predicts a series of \emph{massive} Kaluza-Klein (KK)
graviton resonances with the lightest graviton about 1 TeV.

After the replacement
\begin{equation}\label{kappa_replacement}
\kappa \rightarrow -\kappa \;,
\end{equation}
the RS1 metric becomes \cite{Giudice:05}
\begin{equation}\label{RS_metric_mod}
ds^2 = e^{2 \kappa |y|} \, \eta_{\mu \nu} \, dx^{\mu} \, dx^{\nu} -
dy^2 \;.
\end{equation}
The hierarchy relation is modified as follows
\begin{equation}\label{RS_hierarchy_relation_mod}
\bar{M}_{\mathrm{Pl}}^2 = \frac{\bar{M}_5^3}{\kappa} \left(e^{2 \pi
\kappa r_c} - 1 \right) \;.
\end{equation}
The warp factor $F = \exp(2\kappa|y|)$ and brane cosmological terms
acquire the meanings
\begin{equation}\label{warp_factors_RS_mod}
F\Big|_{y = 0} = 1 \;, \quad F\Big|_{y = \pi r_c} = e^{2\pi\kappa
r_c} \;,
\end{equation}
\begin{equation}\label{tensions_RS_mod}
\Lambda_1 < 0 \;, \quad \Lambda_2
> 0 \;.
\end{equation}

Thus, after replacement \eqref{kappa_replacement} the branes
\emph{are interchanged}. The brane $y = 0$ becomes the TeV brane on
which all the SM fields live. The ratio of the warp factors remains
\emph{the same} (namely, $\exp(2\pi \kappa r_c)$), but the very
value of the warp factor on the TeV brane becomes equal to
\emph{unity}.

In such a scheme, $\bar{M}_5$ can be one to tens TeV, while the
curvature $\kappa$ is allowed to vary from hundred MeV to tens GeV
\cite{Giudice:05,Kisselev:05}. The masses of the KK gravitons are
proportional to the curvature $\kappa$ (see below Eq.
\eqref{graviton_masses}).

Note that after changing variables in \eqref{RS_metric}
\begin{equation}\label{var_change}
x^\mu \rightarrow x'^\mu = e^{- \pi \kappa r_c} \, x^\mu  \;,
\end{equation}
one obtains the metric \cite{Kisselev:05,Kisselev:06}
\begin{equation}\label{RS_metric_change_var}
\quad ds^2 = e^{2 \kappa (\pi r_c - |y|)} \eta_{\mu \nu} \,
dx'^{\mu} \, dx'^{\nu} - dy^2 \;,
\end{equation}
which also leads to the modified hierarchy relation
\eqref{RS_hierarchy_relation_mod}.

The metric with the small curvature, which takes into account the
orbifold structure of the space-time, looks like (for details, see
Appendix~A)
\begin{equation}\label{metric_RSSC}
\mathrm{RSSC}: \quad ds^2 = e^{-2 \sigma(y)} \eta_{\mu \nu} \,
dx^{\mu} \, dx^{\nu} - dy^2 \;,
\end{equation}
with
\begin{equation}\label{sigma_RSSC}
\sigma(y) = \frac{\kappa}{2} \, ( |y| - |\pi r_c - y| - \pi r_c) \;,
\end{equation}
Note that the metrics \eqref{RS_metric_change_var} and
\eqref{metric_RSSC} coincide for $0 < y < \pi r_c$.

The warp factor $F = \exp[-2\sigma(y)]$ with the RSSC function
$\sigma(y)$ \eqref{sigma_RSSC} has the following values at the fixed
points
\begin{equation}\label{F_RSSC}
\mathrm{RSSC}: \quad F\Big|_{y = 0} =  e^{2\pi \kappa r_c} \;, \quad
F\Big|_{y = \pi r_c} = 1  \;.
\end{equation}
The boundary cosmological terms are
\begin{equation}\label{tensions_RSSC}
\mathrm{RSSC}: \quad \Lambda_1 > 0 \;, \quad \Lambda_2 < 0 \;.
\end{equation}
The hierarchy relation,
\begin{equation}\label{RSSC_hierarchy_relation}
\mathrm{RSSC}: \quad  \bar{M}_{\mathrm{Pl}}^2 =
\frac{\bar{M}_5^3}{\kappa} \left(e^{2 \pi \kappa r_c} - 1 \right)
\;,
\end{equation}
is the same as Eq.~\eqref{RS_hierarchy_relation_mod}. In order for
relation \eqref{RSSC_hierarchy_relation} to be satisfied, it is
enough to take $\kappa r_c \approx 10$.%
\footnote{The exact value depends on the ratio $\bar{M}_5^3/\kappa$.
In particular, for $\bar{M}_5 = 1$ TeV, $\kappa = 1$ GeV (100 MeV),
one has $\kappa r_c \simeq 10.2$ (9.8) and $r_c \simeq 2.0$ fm (19.4
fm).}

In between the branes, the 5-dimensional scalar curvature is
negative,
\begin{equation}\label{RS_Ricci_scalar}
\mathcal{R}_5 = -20 \kappa^2 \;,
\end{equation}
while the radius of the curvature is equal to $\kappa^{-1}$ (see
Appendix~A).

Let us use a linear expansion of the metric about its Minkowski
value (see, for instance, \cite{Davoudiasl:00})
\begin{equation}\label{metric_expansion_1}
G_{\mu\nu} = e^{-2 \sigma} \! \bigg( \eta_{\mu\nu} +
\frac{1}{M_5^{3/2}} \, H_{\mu\nu} \bigg) \;, \quad  G_{44} = - 1 +
H_{44} \;.
\end{equation}
After redefinition $H_{\mu\nu} = h_{\mu\nu} + H_{44}/2$, where
$H_{44} = 2\,e^{2\sigma} \phi(x)$ \cite{Toharia:04}, and imposing
transverse-traceless gauge,
\begin{equation}\label{gauge_fixing}
\partial^\mu h_{\mu\nu} = 0 \;, \quad h_\mu^\mu = 0\;,
\end{equation}
the metric fluctuation $h_{\mu\nu}$ describes the massive spin-2
field with 5 degrees of freedom. It can
be decomposed into KK graviton excitations%
\footnote{Due to the form of zero mode wave function, the field
$h^{(0)}_{\mu\nu}$ has only 2 degrees of freedom and describes the
massless graviton \cite{Boos:02}.}
\begin{equation}\label{metric_expansion}
h_{\mu\nu}(x,y) = \frac{1}{\sqrt{2\pi r_c}}\sum_{n=0}^{\infty}
h_{\mu\nu}^{(n)}(x) \, \psi^{(n)}(y) \;.
\end{equation}
The wave functions of the KK excitations obey the equation
\cite{Davoudiasl:00}
\begin{equation}\label{eigenfunc_eq}
\frac{d}{dy} \! \left( e^{-4 \sigma(y)} \frac{d}{dy}\right)
\psi^{(n)}(y) = - m_n^2 \, e^{-2 \sigma(y)} \, \psi^{(n)}(y) \;,
\end{equation}
with $m_n$ being the mass of four-dimensional gravitons,
\begin{equation}\label{four_dim_eq}
(\eta^{\mu\nu}\partial_\mu \partial_\nu + m_n^2) h_{\mu\nu}^{(n)}(x)
= 0 \;.
\end{equation}
Equation \eqref{eigenfunc_eq} follows from 4-dimensional components
of the Einstein-Hilbert equation, if one keeps only linear terms in
$h_{\mu\nu}$. Note that Eq.~\eqref{eigenfunc_eq} means that
\begin{equation}\label{wave_eq}
\square_5 h_{\mu\nu}(x,y) = 0 \;,
\end{equation}
where $\square_5$ is the 5-dimensional d'Alembertian in the
background metric \eqref{metric_RSSC}.

The eigenfunctions $\psi^{(n)}(y)$ satisfy the boundary conditions
\begin{align}
& \frac{d \psi^{(n)}}{dy}\Big|_{y = 0} = 0 \;,
\label{boundary_condition_l} \\
& \frac{d \psi^{(n)}}{dy}\Big|_{y = \pi r_c} = 0
\label{boundary_condition_r} \;,
\end{align}
as well as the orthonormality condition
\begin{equation}\label{orthonormality}
\frac{1}{\pi r_c} \int_0^{\pi r_c} \!\!\! dy \, e^{-2 \sigma(y)}
\,\psi^{(n)}(y) \, \psi^{(m)}(y) = \delta_{nm} \;.
\end{equation}
Thus, we have the Sturm-Liouville problem \eqref{eigenfunc_eq},
\eqref{boundary_condition_l}-\eqref{orthonormality}.

To solve it, let us put for $n>0$
\begin{equation}\label{function_changing}
\psi^{(n)}(y) = z_n^2 \phi^{(n)}(z_n) \;,
\end{equation}
where
\begin{equation}\label{z_n}
z_n = z_n(y) = \frac{m_n}{\kappa} \, e^{\sigma(y)} \;.
\end{equation}
Then we get from \eqref{eigenfunc_eq}
\begin{align}\label{eigenfunc_eq_full}
&\bigg[ z_n^2 (\phi^{(n)}(z_n))'' + z_n (\phi^{(n)}(z_n))' + (z_n^2
- 4) \phi^{(n)}(z_n) \bigg] \left( \frac{\sigma'}{\kappa} \right)^2 \nonumber \\
+ &\bigg[ z_n (\phi^{(n)}(z_n))' + 2 \phi^{(n)}(z_n) \bigg]
\frac{\sigma''}{\kappa^2} = 0 \;.
\end{align}
Here and in what follows the \emph{prime} denotes the derivative
with respect to variable $y$.

The solution of Eq.~\eqref{eigenfunc_eq_full} which satisfies the
\emph{right} boundary condition \eqref{boundary_condition_r} is
given in terms of Bessel functions
\begin{equation}\label{egenfunctions}
\phi^{(n)}(y) = C_n [ J_2 (z_n) Y_1(b_n) - Y_2 (z_n) J_1(b_n) ] \;,
\end{equation}
where
\begin{equation}\label{b_n}
b_n = \frac{m_n}{\kappa} \;,
\end{equation}
and $C_n$ is a constant.

The \emph{left} boundary condition \eqref{boundary_condition_l}
defines the masses of the KK gravitons with respect to the TeV
brane. Taking into account that $\sigma(0) = - \kappa \pi r_c$, we
get
\begin{equation}\label{masses_eq}
J_1 (a_n) Y_1(b_n) - Y_1 (a_n) J_1(b_n) = 0 \;,
\end{equation}
where
\begin{equation}\label{a_n}
a_n = \frac{m_n}{\kappa} \, e^{- \kappa \pi r_c} \;.
\end{equation}

Let us demonstrate that the second term in \eqref{eigenfunc_eq_full}
is equal to zero. Indeed, by using relation $x Z_2'(x) + 2 Z_2(x) =
x Z_1(x)$, where $Z_\nu = J_\nu$ or $Y_\nu$, it can be presented in
the form
\begin{equation}\label{singular_term_eq}
\frac{ C_n}{\kappa} \, [J_1 (z_n) Y_1(b_n) - Y_1 (z_n) J_1(b_n)]
[\delta(y) - \delta(y - \pi r_c)] \;.
\end{equation}
Note that
\begin{equation}\label{z_n_values}
z_n(y) =
\left\{
  \begin{array}{ll}
    a_n, & y = 0 \\
    b_n, & y = \pi r_c
  \end{array}
\right.
\end{equation}
Thus, the expression in \eqref{singular_term_eq} vanishes due to the
boundary condition \eqref{masses_eq}.

As a result, we obtain for $n>0$
\begin{equation}\label{egenfunctions_full}
\psi^{(n)}(y) = N_n e^{2\sigma} \, [ J_2 (z_n) Y_1(b_n) - Y_2 (z_n)
J_1(b_n) ] \;,
\end{equation}
where the normalization constant $N_n$ is defined from the
orthonormality condition \eqref{orthonormality}. An explicit form of
$N_n$ is derived in Appendix~B.

As for the zero mode excitation, its wave function looks like
\begin{equation}\label{zero_wave_func}
\psi^{(0)}(y) = N_0 = \left( \frac{2\pi \kappa r_c}{e^{2\pi\kappa
r_c} - 1} \right)^{\!1/2} = \sqrt{2\pi r_c} \,\,
\frac{\bar{M}_5^{3/2}}{\bar{M}_{\mathrm{Pl}}} \;.
\end{equation}
The orthogonality of the zero mode \eqref{zero_wave_func} and KK
modes \eqref{egenfunctions_full} comes from the equation ($n > 0$)
\begin{align}\label{orthogonality}
\int_0^{\pi r_c} \!\!\!  &dy \, e^{-2 \sigma(y)} \,\psi^{(0)}(y) \,
\psi^{(n)}(y) = \frac{N_0 N_n}{\kappa} \int_{a_n}^{b_n} \frac{dz}{z}
\, [ J_2 (z) Y_1(b_n) - Y_2 (z) J_1(b_n) ]
\nonumber \\
&= \frac{N_0 N_n}{a_n \kappa} [J_1 (a_n) Y_1(b_n) - Y_1 (a_n)
J_1(b_n)] = 0 \;.
\end{align}

The interactions of massless gravitons on the TeV brane are given by
the Lagrangian
\begin{align}\label{massless_lagrangian}
\mathcal{L}_{\mathrm{int}}^{(0)} &= - \frac{1}{\bar{M}_5^{3/2}} \int
\!\! dy \, h_{\mu\nu}^{(0)}(x,y) \, T^{\mu\nu}
(x) \delta (y - \pi \kappa r_c) \nonumber \\
&= - \frac{1}{\bar{M}_{\mathrm{Pl}}} h_{\mu\nu}^{(0)}(x) \,
T^{\mu\nu}(x) \;,
\end{align}
where $T^{\mu \nu}(x)$ is the energy-momentum tensor of the SM
fields.

Let us consider Eq. \eqref{masses_eq} in more detail. Since
\begin{equation}\label{exponetion}
\frac{e^{- \kappa \pi r_c}}{\kappa} \simeq
\frac{1}{\bar{M}_{\mathrm{Pl}}} \left( \frac{\bar{M}_5}{\kappa}
\right)^{3/2} \! \ll 1 \;,
\end{equation}
$m_n$ are defined by the equation
\begin{equation}\label{masses_appr_eq}
J_1(b_n) = 0 \;.
\end{equation}
As a result, the graviton masses have appeared to be proportional to
$\kappa$,
\begin{equation}\label{graviton_masses}
m_n = x_n \kappa \;, \quad n=1,2, \ldots \;,
\end{equation}
where $x_n$ are zeros of the Bessel function $J_1(x)$.

In the limit of a very small curvature,
\begin{equation}\label{small_kappa}
2 \pi\kappa r_c \ll 1 \;,
\end{equation}
one can use asymptotic values of the Bessel functions
\begin{align}\label{Bessel_asymptotic}
J_1 (z) &= \sqrt{\frac{2}{\pi z}} \left[ \sin \left( z - \frac{\pi}{4} \right)
+ \mathrm{O} (|z|^{-1}) \right] \;, \nonumber \\
Y_1 (z) &= - \sqrt{\frac{2}{\pi z}} \left[ \cos \left( z -
\frac{\pi}{4} \right) + \mathrm{O} (|z|^{-1}) \right] \;.
\end{align}
Then we get from \eqref{masses_eq}
\begin{equation}\label{}
\sin (b_n - a_n) \simeq \, \sin(\pi r_c m_n) = 0 \;,
\end{equation}
that results in the well-known graviton spectrum in the model with
one \emph{flat} ED \cite{Arkani-Hamed:98}-\cite{Arkani-Hamed:99}
\begin{equation}\label{masses_ADD}
m_n = \frac{n}{r_c} \;, \quad n = 1, 2, \ldots \;.
\end{equation}
As one can see, $a_n, b_n \simeq n/(\kappa r_c) \gg 1$. Thus, using
asymptotic expressions \eqref{Bessel_asymptotic} was fully
justified.

Let us stress, however, that the AdS$_{\,5}$ space becomes
indistinguishable from a five-dimensional flat space only for
\emph{negligible} values of the curvature $\kappa$. Indeed, in the
limit \eqref{small_kappa}, Eq.~\eqref{RSSC_hierarchy_relation}
transforms into the hierarchy relation for the flat ED
\begin{equation}\label{hierarchy_limit}
\bar{M}_{\mathrm{Pl}}^2 = \bar{M}_5^3 (2\pi r_c) = \bar{M}_5^3 \,
V_1 \;,
\end{equation}
where $V_1$ is the volume of the compact ED. Then the inequality
$2\pi r_c \ll \kappa^{-1}$ means
\begin{equation}\label{kappa_bound}
\kappa \ll \frac{\bar{M}_5^3}{\bar{M}_{\mathrm{Pl}}^2} \simeq 0.17 \cdot 10^{-18}
\left( \frac{\bar{M}_5}{1 \, \mathrm{TeV}} \right)^3 \mathrm{eV} \;.
\end{equation}

The Newton potential between two test masses in the RSSC model was
estimated in \cite{Kisselev:diphotons}
\begin{equation}\label{Newton_potential}
 V(r) = G_N \, \frac{m_1 m_2}{r} \left( 1 + \frac{e^{-m_1 r}}{\pi \kappa r}
\right) \;,
\end{equation}
where $m_1 = x_1 \kappa$ is the mass of the lightest KK graviton,
$x_1 = 3.84$ being the first zero of the Bessel function $J_1(x)$.
Thus, relative corrections to the Newton law appear to be negligible
\cite{Kisselev:diphotons}.

The interaction of the massive KK gravitons with the the SM fields
on the TeV brane is described by the Lagrangian
\cite{Giudice:05}-\cite{Boos:02} (see also \cite{Kisselev:06})
\begin{align}\label{Lagrangian}
\mathcal{L}_{\mathrm{TeV}} &= - \frac{1}{\bar{M}_5^{3/2}}
\sum_{n=1}^{\infty} \int \!\! dy \, \sqrt{G} \,
h_{\mu\nu}^{(n)}(x,y) \, T_{\alpha\beta}(x) \, g^{\mu\alpha}
g^{\nu\beta} \delta (y - \pi \kappa r_c) \nonumber \\
& = - \frac{1}{\Lambda_{\pi}} \, T_{\alpha\beta} (x)
\sum_{n=1}^{\infty} h^{(n)}_{\mu \nu} (x) \, \eta^{\mu\alpha}
\eta^{\nu\beta}\;.
\end{align}
The parameter
\begin{equation}\label{lambda_pi_full}
\Lambda_{\pi} = M_{\mathrm{Pl}} \, \sqrt{ \frac{1 -
Y_1^2(b_n)/Y_1^2(a_n)}{e^{2\kappa \pi r_c} - 1} } \simeq
\bar{M}_{\mathrm{Pl}} \, e^{-\pi \kappa r_c}
\end{equation}
has the meaning of the physical scale on the TeV brane.

In a number of papers (see, for instance, Refs. \cite{Boos:02},
\cite{Rubakov:01}), the linear expansion about the \emph{background}
metric,
\begin{equation}\label{metric_expansion_2}
G_{\mu\nu} = e^{-2 \sigma} \eta_{\mu\nu} + \frac{1}{M_5^{3/2}} \,
\tilde{H}_{\mu\nu}  \;, \quad  G_{44} = - 1 + H_{44} \;.
\end{equation}
is used instead of expansion \eqref{metric_expansion_1}. In such a
case, $\tilde{h}_{\mu\nu} = e^{-2 \sigma} h_{\mu\nu}$, and the
eigenvalue functions are equal to
\begin{align}
\tilde{\psi}^{(0)}(y) &= N_0 \, e^{-2\sigma(y)} \;,
\label{zero_egenfunction} \\
\tilde{\psi}^{(n)}(y) &= N_n \, [ J_2 (z_n) Y_1(b_n) - Y_2 (z_n)
J_1(b_n) ] \;, \label{nonzero_egenfunctions}
\end{align}
with the constants $N_0$ and $N_n$ defined above. The eigenvalue
functions obey the following equations
\begin{equation}\label{eigenfunc_eq_mod}
\left[ \frac{d^2}{dy^2} - 4\sigma'^{\,2}(y) + 2\sigma''(y) \right]
\tilde{\psi}^{(n)}(y) = - m_n^2 e^{2\sigma(y)} \tilde{\psi}^{(n)}(y)
\end{equation}
and boundary conditions
\begin{equation}\label{boundary_conditions_mod}
\left[\frac{d}{dy} + 2\sigma'(y)\right] \tilde{\psi}^{(n)}(y) = 0
\;, \quad \mathrm{for \ } y = 0, \, \pi r_c \;.
\end{equation}
Correspondingly, the orthonormality condition looks like
\begin{equation}\label{orthonormality_mod}
\frac{1}{\pi r_c} \int_0^{\pi r_c} \!\!\! dy \, e^{2 \sigma(y)} \,
\tilde{\psi}^{(n)}(y) \, \tilde{\psi}^{(m)}(y) = \delta_{nm} \;.
\end{equation}

It is clear that the KK graviton masses are defined by
Eq.~\eqref{graviton_masses} as before. Since
$e^{-2\sigma(y)}|_{y=\pi r_c} = 1$, the Lagrangian on the TeV brane
also remains the same for zero mode \eqref{massless_lagrangian} and
massive modes \eqref{Lagrangian}.

\section{Graviton contribution to dilepton production at the LHC}
\label{sec:2}

The goal of this section is to estimate gravity effects in the
dilepton production ($l = e$ or $\mu$),
\begin{equation}\label{process}
p \, p \rightarrow l^+ l^- + X \;,
\end{equation}
at the LHC  in the RSSC model. The formulas for the $p_{\perp}$
distribution of the leptons are presented in Appendix~C. At fixed
values of the dimensionless variable $x_{\perp} =
2p_{\perp}/\sqrt{s}$, the gravity cross section has the following
dependence on fundamental gravity scale $\bar{M}_5$
\begin{equation}\label{gravity_cs_1}
\frac{d \sigma (\mathrm{grav})}{d p_{\perp}} \sim
\frac{1}{\bar{M}_5^3} \;.
\end{equation}

For numerical calculations, we impose the cut on the lepton
pseudorapidity used by the CMS Collaboration. For the dimuon events
it looks like
\begin{equation}\label{rapidity_cut_mu}
|\eta| < 2.4 \;,
\end{equation}
while for the dielectron events the cuts are the following%
\footnote{The transition region $1.44 < |\eta| < 1.57$ ($1.37 <
|\eta| < 1.52$)  between the ECAL barrel and endcap calorimeters is
usually excluded in the CMS (ATLAS) experiment.}
\begin{equation}\label{rapidity_cut_el}
|\eta| < 1.44 \;, \quad  1.57 < |\eta| < 2.50 \;.
\end{equation}
The reconstruction efficiency $85 \%$ is assumed for the dilepton
events \cite{CMS_dilepton_efficiency}.

We use the MSTW NNLO parton distributions~\cite{MSTW}, and convolute
them with the partonic cross sections. The PDF scale is taken to be
equal to the invariant mass of the lepton pair, $\mu = M_{l^+l^-}$.
In order to take into account SM higher order corrections, the $K$
factor 1.5 is used for the SM background, while a conservative value
of $K=1$ is taken for the signal.

The differential cross section of the process under consideration
has three terms
\begin{equation}\label{cross_sec_sum}
d\sigma = d \sigma (\mathrm{SM}) + d\sigma (\mathrm{grav}) + d\sigma
(\mathrm{SM}\mathrm{-grav}) \;,
\end{equation}
where the last one comes from the interference between the SM and
graviton interactions. Since the SM amplitude is pure real, while
the real part of each graviton resonance is antisymmetric with
respect to its central point, the interference term has appeared to
be negligible in comparison with the pure gravity and SM terms after
integration in partonic momenta~\cite{Kisselev:diphotons}.

The account of the \emph{graviton widths} is a crucial point for
both analytical calculations and numerical estimations. As it was
shown in our previous papers~\cite{Kisselev:diphotons},
\cite{Kisselev:13}, an ignorance of the graviton widths is a
\emph{rough} approximation, since it results in very large
suppression of the cross sections. The reason lies partially in the
fact that
\begin{equation}\label{gravity_cs_2}
\frac{d \sigma (\mathrm{grav})}{d p_{\perp}} \sim
\frac{1}{p_{\perp}^3} \left( \frac{\sqrt{s}}{\bar{M}_5} \right)^3
\;,
\end{equation}
while in zero width approximation one gets
\begin{equation}\label{gravity_cs_zero}
\left. \frac{d \sigma (\mathrm{grav})}{d p_{\perp}}
\right|_{\mathrm{zero \ widths}} \sim \frac{1}{\bar{M}_5^3} \left(
\frac{\sqrt{s}}{\bar{M}_5} \right)^3 \;.
\end{equation}

Let us stress that in the RSSC model the gravity cross sections
\emph{do not depend} on the curvature $\kappa$ (up to small power
corrections), provided $\kappa \ll \bar{M}_5$, in contrast to the
standard RS1 model in which all bounds on $\bar{M}_5$ depend on the
ratio $\kappa/\bar{M}_{\mathrm{Pl}}$~\cite{Randall:99}.

\begin{figure}[hbtp]
 \begin{center}
 \resizebox{7cm}{!}{\includegraphics{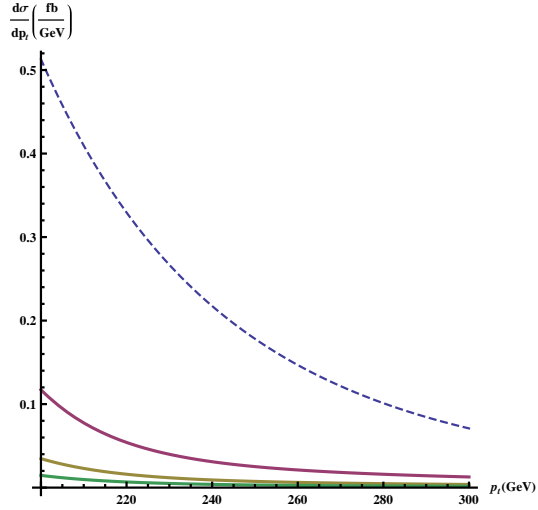}}
 \caption{The KK graviton contribution to the dielectron production for $\bar{M}_5 = 2, 4, 6$ TeV
 (solid curves, from above) vs. SM (Born) contribution (dashed curve) at $\sqrt{s} = 8$ TeV.}
 \label{fig:cs_1_8TeV}
 \end{center}
\end{figure}

\begin{figure}[hbtp]
 \begin{center}
 \resizebox{7cm}{!}{\includegraphics{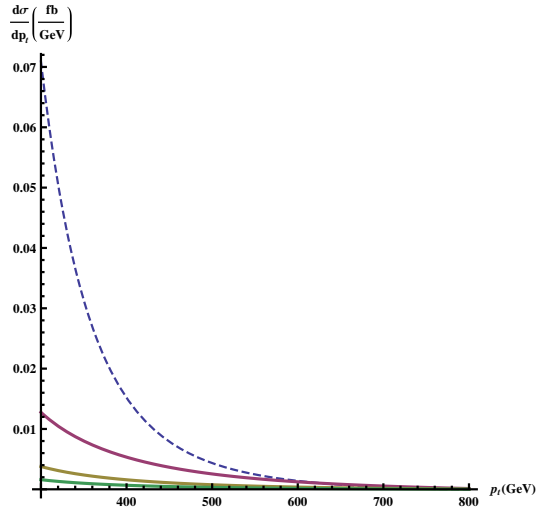}}
 \caption{The same as in Fig.~\ref{fig:cs_1_8TeV}, but for larger values of $p_{\perp}$.}
    \label{fig:cs_3_8TeV}
  \end{center}
\end{figure}

\begin{figure}[hbtp]
 \begin{center}
 \resizebox{7cm}{!}{\includegraphics{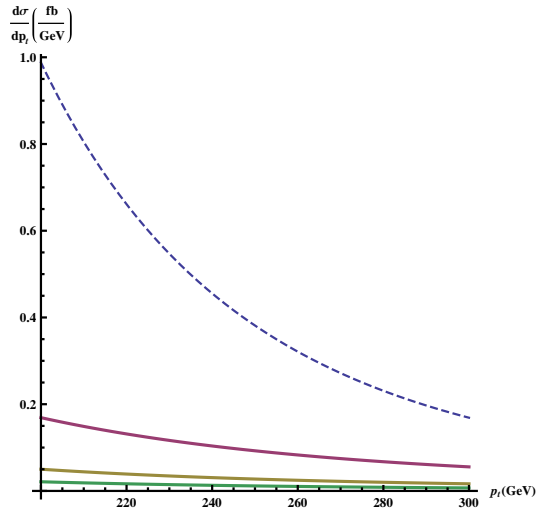}}
 \caption{The KK graviton contribution to the dielectron production for $\bar{M}_5 = 4, 6, 8$ TeV
 (solid curves, from above) vs. SM (Born) contribution (dashed curve) at $\sqrt{s} = 13$ TeV.}
 \label{fig:cs_1_13TeV}
 \end{center}
\end{figure}

\begin{figure}[hbtp]
 \begin{center}
 \resizebox{7cm}{!}{\includegraphics{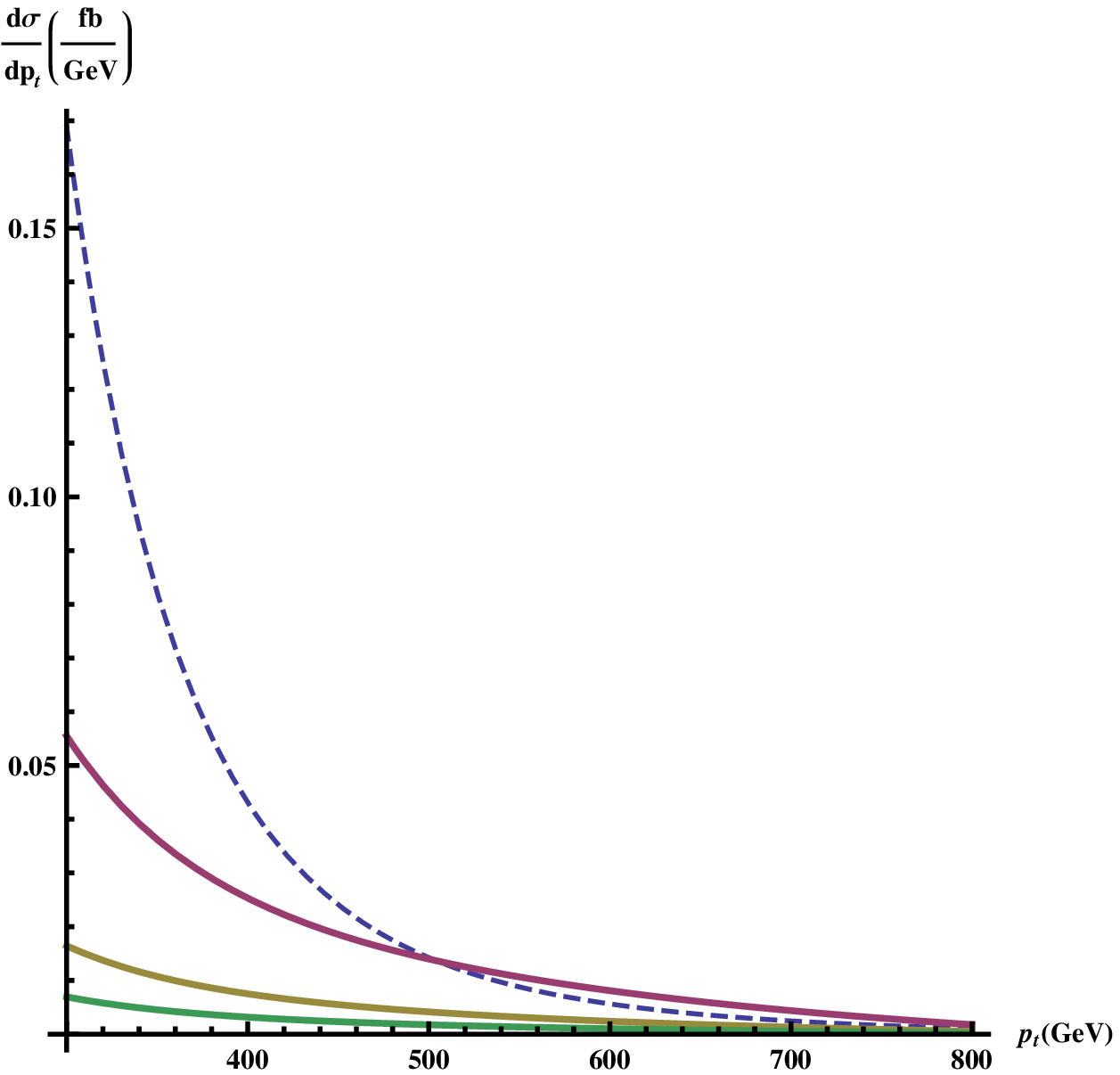}}
 \caption{The same as in Fig.~\ref{fig:cs_1_13TeV}, but for larger values of $p_{\perp}$.}
    \label{fig:cs_3_13TeV}
  \end{center}
\end{figure}

In Figs.~\ref{fig:cs_1_8TeV} and \ref{fig:cs_3_8TeV} we present the
gravity cross sections for the dielectron events at 8 TeV LHC.
The differential cross sections at 13 TeV are shown in
Figs.~\ref{fig:cs_1_13TeV} and \ref{fig:cs_3_13TeV}. Note that the
gravity mediated contributions to the cross sections do not include
the SM contribution (i.e. solid lines in all figures correspond to
pure gravity contributions).

The figures for dimuon cross sections look similar to Figs.
\ref{fig:cs_1_8TeV}-\ref{fig:cs_3_13TeV}.


Let $N_S$($N_B$) be a number of signal (background) dilepton events
with $p_{\perp} > p_{\perp}^{\mathrm{cut}}$,
\begin{equation}\label{ev_numder}
N_S  = \!\! \int\limits_{p_{\perp} > p_{\perp}^{\mathrm{cut}}} \!\!
\frac{d \sigma (\mathrm{grav})}{dp_{\perp}} \, dp_{\perp}  \;, \quad
N_B = \!\! \int\limits_{p_{\perp} > p_{\perp}^{\mathrm{cut}}} \!\!
\frac{d \sigma (\mathrm{SM})}{dp_{\perp}} \, dp_{\perp}
 \;.
\end{equation}

\begin{figure}[hbtp]
 \begin{center}
 \resizebox{7cm}{!}{\includegraphics{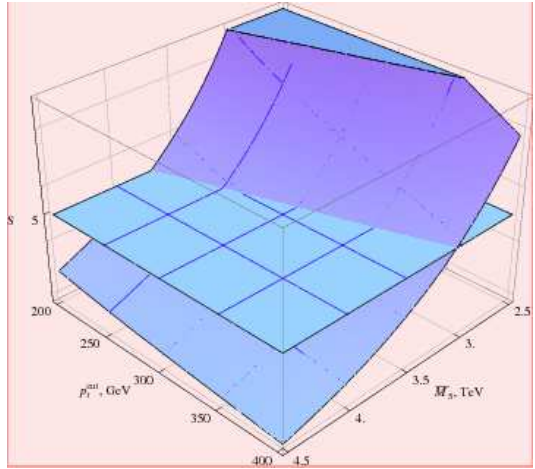}}
 \caption{The statistical significance $S$ for the dilepton ($\mathrm{\mu + e}$ )
 production at the LHC for $\sqrt{s} = (7+8)$ TeV and integrated
 luminosity (5+20) fb$^{-1}$ as a function of the transverse momentum cut
 $p_{\perp}^{\mathrm{cut}}$ and \emph{reduced} 5-dimensional gravity
 scale $\bar{M}_5$. The plane $\mathcal{S}=5$ is also shown.}
 \label{fig:S_7_8TeV}
 \end{center}
\end{figure}

\begin{figure}[hbtp]
 \begin{center}
 \resizebox{7cm}{!}{\includegraphics{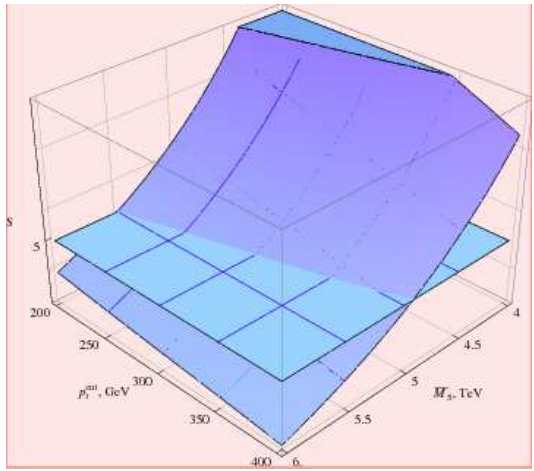}}
 \caption{The same as in figure \ref{fig:S_7_8TeV}, but for
 $\sqrt{s} = 13$ TeV and integrated luminosity 30 fb$^{-1}$.}
 \label{fig:S_13TeV}
 \end{center}
\end{figure}

Then we define the statistical significance $\mathcal{S} =
N_S/\sqrt{N_B + N_S}$, and require a $5 \sigma$ effect. In
Fig.~\ref{fig:S_7_8TeV} the statistical significance is shown for
total number of ``events'' with $\sqrt{s} = 7$ TeV and $\sqrt{s} =
8$ TeV as a function of the transverse momentum cut
$p_{\perp}^{\mathrm{cut}}$ and \emph{reduced} 5-dimensional gravity
scale $\bar{M}_5$. The integrated luminosity was taken to be 5
fb$^{-1}$ and 20 fb$^{-1}$ for $\sqrt{s} = 7$ TeV and $\sqrt{s} = 8$
TeV, respectively.
Figure \ref{fig:S_13TeV} represents the significance $\mathcal{S}$
for the dilepton events with $\sqrt{s} = 13$ TeV and 30 fb$^{-1}$.

Previously, calculations of dilepton cross sections were done in
\cite{Giudice:05} \emph{without} taking into account finite widths
of the KK gravitons. As was shown in \cite{Kisselev:13} (see also
\cite{Kisselev:diphotons}), in zero width approximation the gravity
cross sections are very small in comparison with the background
cross section at low and moderate values of $p_{\perp}$. That is
why, a high cut $p_{\perp}^{\mathrm{cut}}$ is needed in order to get
$N_S$ comparable with $N_B$. Correspondingly, LHC search limits have
appeared to be significantly smaller than in our case.



\section{Conclusions}
\label{sec:3}

In the present paper the RSSC
model~\cite{Kisselev:05}-\cite{Kisselev:06} is considered in detail
in which the reduced 5-dimensional Planck scale $\bar{M}_5$ can vary
from few TeV to tens TeV, while the curvature $\kappa$ is allowed to
vary from hundred MeV to few GeV. In fact, the only condition
$\kappa \ll \bar{M}_5$ should be satisfied. The mass spectrum and
experimental signature of the model with the small curvature are
similar to those in the ADD model~\cite{Arkani-Hamed:98} with one
flat extra dimension.

The $p_{\perp}$-distributions for the lepton pairs with high
$p_{\perp}$ at the LHC are calculated for the collision energies 7,
8, and 13 TeV (for the dielectron case, the results of our
calculations are shown in
Figs.~\ref{fig:cs_1_8TeV}-\ref{fig:cs_3_8TeV} and
Figs.~\ref{fig:cs_1_13TeV}-\ref{fig:cs_3_13TeV}).%
\footnote{To reduce a number of figures, we did not present cross
sections for the energy $\sqrt{s} = 7$ TeV and cross sections for
the dimuon events, since they look very similar to the dielectron
cross sections.}

The account of the KK graviton widths is the crucial point for the
numerical calculations, since the zero width approximation
significantly underestimates the gravity cross sections even at
large $p_{\perp}$.

The statistical significance as a function of the \emph{reduced}
5-dimensional Planck scale $\bar{M}_5$ and cut on the lepton
transverse momentum $p_{\perp}^{\mathrm{cut}}$ is calculated (see
Figs.~\ref{fig:S_7_8TeV}-\ref{fig:S_13TeV}). Let us underline that
\emph{both the dielectron and dimuon} events are taken into account.

Recently, a search for large extra spatial dimensions in the dimuon
(dielectron) channel was presented using a data sample of 20.6
fb$^{-1}$ (19.6 fb$^{-1}$) at the center-of-mass energy of 8 TeV
collected by the CMS detector
\cite{CMS:dimuons}-\cite{CMS:dielectrons}. No significant deviations
from SM were observed. Previously, dilepton spectra were found to be
consistent with SM expectations at the energy of 7 TeV
\cite{CMS:dileptons}.

By using our calculations for 7 and 8 TeV, we conclude that in the
RSSC framework the region
\begin{equation}\label{search limit_7+8_TeV}
M_5 < 6.84 \mathrm{\ TeV}
\end{equation}
is excluded at 95$\%$ C.L. Note that for the ADD scenario the
exclusion limits on the model parameter $M_S$ are 4.49 and 4.77 TeV
for the dimuon and dielectron events, respectively
\cite{CMS:dimuons}-\cite{CMS:dielectrons}.

Correspondingly, we obtain the discovery limit for the 13 TeV LHC
with the integrated luminosity 30 fb$^{-1}$:
\begin{equation}\label{search limit_13_TeV}
M_5 = 10.16 \mathrm{\ TeV} \;.
\end{equation}
In deriving Eqs.~\eqref{search limit_7+8_TeV}, \eqref{search
limit_13_TeV}, we used the relation $M_5 = (2 \pi)^{1/3} \bar{M}_5$
(\ref{grav_scale_reduced}) in order to present the bounds on
fundamental gravity scale $M_5$.

It is important that these bounds on $M_5$ do not depend on the
curvature $\kappa$ (up to small powerlike corrections), contrary to
the original RS1 model~\cite{Randall:99} in which estimated bounds
on $M_5$ depend on the ratio $\kappa/\bar{M}_{\mathrm{Pl}}$.

Previously, analogous bounds were obtained for the diphoton
production~\cite{Kisselev:diphotons}. Recently, the LHC search
limits were estimated for dimuon events~\cite{Kisselev:13}. Dilepton
production at very high luminosities (HL-LHC) was studied in
\cite{Kisselev:12}.



\section*{Acknowledgements}

The author is indebted to V.A.~Petrov for useful discussions.



\setcounter{equation}{0}
\renewcommand{\theequation}{A.\arabic{equation}}

\section*{A Warped metric of RSSC model}
\label{app:A}

From the action \eqref{RS_action}, 5-dimensional Einstein-Hilbert's
equations follow
\begin{align}\label{H-E_equation}
\sqrt{|G|} & \left( \mathcal{R}_{MN} - \frac{1}{2} \, G_{MN}
\mathcal{R} \right) = - \frac{1}{2} \Big[ \sqrt{|G|} \, G_{MN}  \Lambda \nonumber \\
&+  \sqrt{|g^{(1)}|} \, g^{(1)}_{\mu\nu} \, \delta_M^\mu \,
\delta_N^\nu \, \delta(y) \, \Lambda_1 +  \sqrt{|g^{(2)}|} \,
g^{(2)}_{\mu\nu} \, \delta_M^\mu \, \delta_N^\nu \, \delta(y - \pi
r_c) \, \Lambda_2 \Big] \;.
\end{align}
In order to solve them, let us assume that the background metric
respects 4-dimensional Poincare invariance ($\mu = 0, 1,2,3$)
\begin{equation}\label{RS_background_metric}
\quad ds^2 = e^{-2 \sigma (y)} \, \eta_{\mu \nu} \, dx^{\mu} \,
dx^{\nu} - dy^2 \;,
\end{equation}
After orbifolding (see Section~\ref{sec:1}), the coordinate of the
extra compact dimension varies within the limits $0 \leqslant  y
\leqslant \pi r_c$.

Let us define
\begin{equation}\label{cov_4dim_metric}
g_{\mu\nu} = e^{-2 \sigma (y)} \, \eta_{\mu\nu} \;.
\end{equation}
Then the 5-dimensional background metric tensor looks like
\begin{equation}\label{cov_metric_tensor}
G_{M\!N} = \left(
  \begin{array}{cc}
  g_{\mu\nu} & 0 \\
    0 & -1 \\
  \end{array}
\right) \;.
\end{equation}
Correspondingly,
\begin{equation}\label{cont_metric_tensor}
G^{M\!N} = \left(
  \begin{array}{cc}
  g^{\mu\nu} & 0 \\
    0 & -1 \\
  \end{array}
\right) \;,
\end{equation}
with
\begin{equation}\label{cont_4dim_metric}
g^{\mu\nu} = e^{2 \sigma (y)} \, \eta^{\mu\nu} \;.
\end{equation}

Non-trivial elements of the Christoffel symbols of the second kind
looks like (there is no summation in $\mu$)
\begin{align}\label{Cristoffel_symbol_elements}
\Gamma^\mu_{\mu 4} &= \frac{1}{2} \, g^{\mu\mu} \, \frac{\partial
g_{\mu\mu}}{\partial y} = - \sigma'(y) \, \delta_\mu^\mu \;, \nonumber \\
\Gamma^4_{\mu \mu} &= \frac{1}{2} \, \frac{\partial
g_{\mu\mu}}{\partial y}  = - \sigma'(y) \, g_{\mu\mu} \;.
\end{align}

As a result, non-zero elements of the curvature tensor
$\mathcal{R}_{K\!M,N\!L}$ are the following ($\mu \neq \nu$)
\begin{align}\label{curvature_tensor_elements}
\mathcal{R}_{\mu 4, 4 \mu} &= \mathcal{R}_{4 \mu, \mu 4} = -
\mathcal{R}_{\mu 4, \mu 4} = - \mathcal{R}_{4 \mu, 4 \mu}  =
[\sigma'^2(y) - \sigma''(y)] \, g_{\mu\mu} \;, \nonumber \\
\mathcal{R}_{\mu \nu, \nu \mu} &= \mathcal{R}_{\nu \mu, \mu \nu} = -
\mathcal{R}_{\mu \nu, \mu \nu} = - \mathcal{R}_{\nu \mu, \nu \mu} =
- \sigma'^2(y)   \, g_{\mu\mu}  \, g_{\nu\nu}\;.
\end{align}
The non-zero elements of the Ricci tensor $\mathcal{R}_{M\!N} =
G^{K\!L} \, \mathcal{R}_{K\!M,N\!L}$ are
\begin{align}\label{bulk_Ricci_tensor}
\mathcal{R}_{44} &=  4[\sigma'^2(y) - \sigma''(y)] \;, \nonumber
\\
\mathcal{R}_{\mu\mu} &=  [- 4\sigma'^2(y) + \sigma''(y)] \,
g_{\mu\mu} \;.
\end{align}
The 5-dimensional scalar curvature $\mathcal{R}_5 = G^{M\!N} \,
\mathcal{R}_{M\!N}$ is equal to
\begin{equation}\label{Ricci_scalar}
\mathcal{R}_5 = - 20 \sigma'^2(y) + 8 \sigma''(y) \;.
\end{equation}

Then the Einstein-Hilbert's equations are reduced to
\begin{align}
\sigma'^2 (y) &= - \frac{\Lambda}{12} \;, \label{RS_sigma_eq_1} \\
\sigma''(y) &= \frac{1}{6} \, [\Lambda_1 \, \delta(y) + \Lambda_2 \,
\delta(\pi r_c - y)] \;. \label{RS_sigma_eq_2}
\end{align}
The first solution of this set of equations was presented in
Ref.~\cite{Randall:99}
\begin{equation}\label{RS_sigma}
\sigma_{\mathrm{RS}}(y) = \kappa \, |y| \;,
\end{equation}
with the cosmological constant
\begin{equation}\label{RS_cosmol_constant}
\Lambda_{\mathrm{RS}} = -12 \kappa^2 \;,
\end{equation}
and boundary cosmological terms
\begin{equation}\label{RS_brane_tensions}
\Lambda_1^{\mathrm{RS}} = - \Lambda_2^{\mathrm{RS}} = 12 \kappa \;.
\end{equation}
Here $\kappa$ is a scale with the dimension of mass.

However, we get from \eqref{RS_sigma} that
$\sigma''_{\mathrm{RS}}(y) = 2 \kappa \, \delta(y)$ instead of
$\sigma''(y) = \kappa [ \delta(y) - \delta(\pi r_c - y) ]$.
Moreover, Eqs.~\eqref{RS_sigma_eq_1}, \eqref{RS_sigma_eq_2} say us
that the cosmological constant $\Lambda$ \emph{should depend} on
coordinate $y$.

In the bulk the set of equations looks like
\begin{align}
\sigma'^2 (y) &=  - \frac{\Lambda}{12} \label{sigma_bulk_eq_1} \\
\sigma''(y) &= 0 \;, \label{sigma_bulk_eq_2}
\end{align}
with the evident solution
\begin{equation}\label{sigma_lambda_bulk}
\sigma(y) = \kappa y + \mathrm{constant}\;, \quad \Lambda = -12
\kappa^2 \;.
\end{equation}

Then we have for the full interval $0 \leqslant y \leqslant \pi r_c$
\begin{align}
\sigma'^2 (y) &= \kappa^2 z(y) \;, \label{RSSC_sigma_eq_1} \\
\sigma''(y) &= \frac{1}{6} \, [\Lambda_1 \, \delta(y) + \Lambda_2 \,
\delta(\pi r_c - y)] \;, \label{RSSC_sigma_eq_2}
\end{align}
where $z(y) = 1$ for $0 < y < \pi r_c$.

One obtains from Eq.~\eqref{RSSC_sigma_eq_2}
\begin{equation}\label{sigma_derivative}
\sigma'(y) = \frac{1}{12} \, [\Lambda_1 \, \tilde{\varepsilon}(y) +
\Lambda_2 \, \tilde{\varepsilon}(y - \pi r_c )] + A \;,
\end{equation}
where $A$ is a constant, and
\begin{equation}\label{epsilon_definition}
\tilde{\varepsilon}(x) = \left\{
  \begin{array}{cc}
    d|x|/dx \;, & |x| > 0  \\
    1 \;, & x = 0
  \end{array}
\right.
\end{equation}
In fact, $\tilde{\varepsilon}(x)$ is the function $\varepsilon (x) =
\theta(x) - \theta (-x)$ supplemented by
its value at $x=0$.%
\footnote{Otherwise, $\Lambda$ will be \emph{uncertain} at the
boundary points $y=0$ and $y = \pi r_c$.}
Let us note that $[\tilde{\varepsilon}(x)]^2 = 1$ for all $x$, and
\begin{equation}\label{epsilon_combination}
\frac{1}{2} \, \left[ \tilde{\varepsilon}(y) +
\tilde{\varepsilon}(\pi r_c - y) \right] = \left\{
  \begin{array}{ll}
    0 \;, & y < 0 \;,  \\
    1 \;, & 0 \leqslant y \leqslant \pi r_c \;, \\
    0 \;, & y > \pi r_c \;.
  \end{array}
\right.
\end{equation}
Since the choice of $A$ is equivalent to a redefinition of
$\Lambda_1$ and $\Lambda_2$, in what follows, we can put $A = 0$.

As for the boundary cosmological terms $\Lambda_{1,2}$ and function
$z(y)$, we get from \eqref{RSSC_sigma_eq_1} the equation
\begin{equation}\label{z_equation}
z(y) = \frac{1}{(12\kappa)^2} \, [\Lambda_1^2 + \Lambda_2^2 - 2
\Lambda_1 \Lambda_2 \, \tilde{\varepsilon}(y) \, \tilde{\varepsilon}
(\pi r_c - y) ] \;.
\end{equation}
Thus, we obtain%
\footnote{We used the relation $1 + \tilde{\varepsilon}(y) \,
\tilde{\varepsilon} (\pi r_c - y) = \tilde{\varepsilon} (y) +
\tilde{\varepsilon} (\pi r_c - y)$ valid for all $y$.}
\begin{align}
\Lambda &= -6\kappa^2 [ \tilde{\varepsilon}(y) +
\tilde{\varepsilon}(\pi r_c -
y)] \;, \label{lambda_solution} \\
\Lambda_1 &= - \Lambda_2 = 6 \kappa \;.
\label{brane_tensions_solutions}
\end{align}
As one can see from Eqs. \eqref{lambda_solution} and
\eqref{epsilon_combination}, the cosmological constant $\Lambda$ is
equal to $(-12\kappa^2)$ at $0 \leqslant y \leqslant \pi r_c$, and
it is zero outside this region.

Finally, we find
\begin{equation}\label{RSSC_sigma_preliminary}
\sigma(y) = \frac{\kappa}{2} \, ( |y| - |\pi r_c - y| ) + B \;,
\end{equation}
where $B$ is a constant. Note that Eqs.~\eqref{lambda_solution},
\eqref{brane_tensions_solutions} differ from RS1 fine tuning
solutions \eqref{RS_cosmol_constant}, \eqref{RS_brane_tensions}.

To get the RSSC scenario, we take%
\footnote{The choice of a particular value of $B$ is equivalent to
changing variables $x^\mu \rightarrow e^B x^\mu$.}
\begin{equation}\label{constant}
B = - \frac{1}{2} \, \pi \kappa \, r_c \;.
\end{equation}
Then
\begin{equation}\label{RSSC_sigma}
\sigma(y) = \frac{\kappa}{2} \, ( |y| - |\pi r_c - y| - \pi r_c) \;,
\end{equation}
and we come to the metric \eqref{metric_RSSC}.

The function $\sigma(y)$ \eqref{RSSC_sigma} is $Z_2$--symmetric due
to the periodicity condition (points $y - \pi r_c$ and $y + \pi r_c$
are identified). The $y$-dependent part of $\sigma(y)$ is symmetric
under substitutions
\begin{equation}\label{sigma_symmetry}
y \rightarrow \pi r_c - y \;, \quad \kappa \rightarrow - \kappa \;.
\end{equation}
It means that the branes are interchanged if we take an opposite
sign for $\kappa$ (see comments to
Eqs.~\eqref{kappa_replacement}-\eqref{tensions_RS_mod} in the main
text).

Note that the the Ricci tensor \eqref{Ricci_scalar} is proportional
to the metric tensor \emph{only} in between the branes,
\begin{equation}\label{Ricci_tensor}
\mathcal{R}_{M\!N} = -4 \kappa^2 G_{M\!N} = \frac{\Lambda}{3} \,
G_{M\!N} \;, \quad 0 < y < \pi r_c\;,
\end{equation}
with $\Lambda$ being cosmological constant. For a space-time with
the constant curvature $K$, the following equation holds
\begin{equation}\label{constant_curvature_eq}
R_{K\!M,N\!L} = K \left( G_{K\!N} \,  G_{M\!L}  - G_{K\!L} \,
G_{M\!N} \right) \;,
\end{equation}
and we find
\begin{equation}\label{constant_curvature}
K = \kappa^2  \;.
\end{equation}
Correspondingly, the radius of the curvature \emph{in the bulk} is
\begin{equation}\label{curvature_radius}
\rho = \frac{1}{\kappa} \;.
\end{equation}

At the boundaries, both the tensors $\mathcal{R}_{K\!M,N\!L}$,
$\mathcal{R}_{M\!N}$ and scalar curvature $\mathcal{R}_5$ have
singular terms, as one can see from
Eqs.~\eqref{curvature_tensor_elements}-\eqref{Ricci_scalar} and
\eqref{RSSC_sigma_eq_2}.



\setcounter{equation}{0}
\renewcommand{\theequation}{B.\arabic{equation}}

\section*{B Normalization of graviton wave functions}
\label{app:B}

The normalization constants are obtained from the following relation
($n>0$)
\begin{equation}\label{normalization}
1 = \frac{1}{\pi r_c} \int_0^{\pi r_c} \!\! dy \, e^{-2\sigma(y)} [
\psi^{(n)}(z_n)]^2  = \frac{1}{\pi\kappa r_c b_n^2} \, N_n^2 \, I_n
\;,
\end{equation}
where
\begin{equation}\label{norm_integral_1}
I_n = \int_{a_n}^{b_n}  \!\! dz z \left[ J_2(z) Y_1(b_n) - Y_2(z)
J_1(b_n) \right]^2 \;.
\end{equation}
The parameters $a_n$ and $b_n$ are defined in the main text (see
Eqs. \eqref{a_n}, \eqref{b_n}).

To calculate $I_n$, we use the table integral \cite{Prudnikov},
\begin{align}\label{Bessel_int_2}
\int^x \!\!\! dz Z_\nu (z) Z^{\,\prime}_\nu (z) = \frac{x^2}{4}
\big[ &2 Z_\nu (x) Z^{\,\prime}_\nu (x) - Z_{\nu+1}(x) Z^{\,\prime}_{\nu-1}(x)
\nonumber \\
&- Z_{\nu-1}(x) Z^{\,\prime}_{\nu +1}(x) \big] + \mathrm{constant}
\;,
\end{align}
where $Z_\nu, \, Z^{\,\prime}_\nu = J_\nu$ or $Y_\nu$, as well as
relations between Bessel functions,
\begin{align}\label{Bessel_relation}
J_{\nu+1}(x)Y_{\nu}(x) - Y_{\nu+1}(x) J_{\nu}(x) &= \frac{2}{\pi x}
\;,
\nonumber \\
J_{\nu+2}(x)Y_{\nu}(x) - Y_{\nu+2}(x) J_{\nu}(x) &=
\frac{4(\nu+1)}{\pi x^2} \;.
\end{align}
Then the integral \eqref{norm_integral_1} is equal to
\begin{align}\label{norm_integral_2}
I_n &= 2a_n [ J_1 (a_n) Y_1(b_n) - Y_1 (a_n) J_1(b_n) ] [ J_2 (a_n)
Y_1(b_n) - Y_2 (a_n) J_1(b_n) ] \;, \nonumber \\
&- \frac{a_n^2}{2} [ J_1 (a_n) Y_1(b_n) - Y_1 (a_n) J_1(b_n) ]^2
\nonumber \\
&- \frac{a_n^2}{2} [ J_2 (a_n) Y_1(b_n) - Y_2 (a_n) J_1(b_n) ]^2 +
\frac{2}{\pi^2} \;.
\end{align}

Due to the left boundary condition in the form~\eqref{masses_eq},
two first terms in the r.h.s of Eq. \eqref{norm_integral_2} vanish,
and we get
\begin{equation}\label{norm_integral_exp}
I_n = \frac{2}{\pi^2} \left[ 1 - \frac{Y_1^2(b_n)}{Y_1^2(a_n)}
\right] =  \frac{2}{\pi^2} \left[ 1 - \frac{J_1^2(b_n)}{J_1^2(a_n)}
\right] \;.
\end{equation}
As a result, we find the expression for the normalization constant
\begin{equation}\label{norm_constant_exp}
N_n^{-2} = \frac{2}{\kappa r_c \pi^3 b_n^2} \left[ \frac{Y_1^2(a_n)
- Y_1^2(b_n)}{Y_1^2(a_n)}\right] \;.
\end{equation}
Our formula \eqref{norm_constant_exp} is in agreement with that from
Ref.~\cite{Giudice:05}.%
\footnote{Note that our notations are somewhat different.}

For $n \neq m$, we have
\begin{equation}\label{orthogonality_n_m}
\int_0^{\pi r_c} \!\!\! dy \, e^{-2 \sigma(y)} \,\psi^{(n)}(y) \,
\psi^{(m)}(y) = \frac{1}{\kappa} \, N_n N_m \, I_{nm} \;,
\end{equation}
where
\begin{align}\label{orthogonal_int}
I_{nm} = \int_{e^{-\pi\kappa r_c}}^{1}  \!\!\! dz z \, &[ J_2 (b_n
z) Y_1(b_n) - Y_2 (b_n z) J_1(b_n) ] \nonumber \\
\times &[ J_2 (b_m z) Y_1(b_m) - Y_2 (b_m z) J_1(b_m)] \;.
\end{align}
By using table integral \cite{Prudnikov},
\begin{align}\label{Bessel_int_3}
\int^x \!\!\! dz z Z_\nu (az) Z^{\,\prime}_\nu (bz) &= \frac{x}{a^2
- b^2} [ a Z_{\nu+1}(ax) Z^{\,\prime}_{\nu}(b x) - b Z_{\nu}(ax)
Z^{\,\prime}_{\nu+1}(b x)] \nonumber
\\
&+ \mathrm{constant} \;,
\end{align}
and Eq.~\eqref{masses_eq}, one can be easily show that $I_{nm} =0$.



\setcounter{equation}{0}
\renewcommand{\theequation}{C.\arabic{equation}}

\section*{C Cross section for dilepton production}
\label{app:C}

The differential cross section of the DY process \eqref{process} is
given by ($l = e$ or $\mu$)
\begin{align}\label{cross_sec}
\frac{d \sigma}{d p_{\perp}}(p p \rightarrow  l^+ l^- + X) &=
2p_{\perp} \!\!\!\! \sum\limits_{a,b = q,\bar{q},g} \!\!
\int\nolimits \!\! \frac{d\tau \sqrt{\tau}}{\sqrt{\tau -
x_{\perp}^2}} \! \int\nolimits \! \frac{dx_1}{x_1}  f_{a/p}(\mu^2,
x_1) \nonumber \\
&\times f_{b/p}(\mu^2, \tau/x_1) \, \frac{d \sigma}{d\hat{t}}(a b
\rightarrow l^+ l^-) \;,
\end{align}
with the transverse energy of the lepton pair equals to
$2p_{\perp}$. In \eqref{cross_sec} two dimensionless quantities are
introduced
\begin{equation}\label{tau_xtr}
x_{\perp} = \frac{2 p_{\perp}}{\sqrt{s}} \;, \quad \tau = x_1 x_2
\,,
\end{equation}
where $x_2$ is the momentum fraction of the parton $b$ in
\eqref{cross_sec}.

Without cuts, integration variables in \eqref{cross_sec} vary within
the following limits
\begin{equation}\label{int_region_full}
x_{\perp}^2 \leq \tau \leq 1 \;, \quad \tau  \leq x_1 \leq 1 \;.
\end{equation}
After imposing kinematical cut on lepton rapidity, the integration
region becomes more complicated (see Appendix~A in
Ref.~\cite{Kisselev:13}).

The SM contribution to the $p_{\perp}$-distribution looks like
\begin{equation}
\frac{d\sigma^{\mathrm{SM}}}{d\hat{t}}(q \bar{q} \rightarrow l^+
l^-) = \frac{1}{48 \pi s^2} \left[ u^2 \left( |G^{LL}|^2 +
|G^{RR}|^2 \right) + t^2 \left( |G^{LR}|^2 + |G^{RL}|^2 \right)
\right] \;,
\end{equation}
with
\begin{equation}
G^{AB}(s) = \sum_{V=\gamma,Z} \! \frac{g_A(V \rightarrow l^+l^-) \,
g_A(V \rightarrow q \bar{q})}{s - m_V^2 + i m_V \Gamma_V} \;.
\end{equation}
Here $g_{L(R)}(\gamma \rightarrow l^+l^-) = g_{L(R)}(\gamma
\rightarrow q \bar{q}) = e$, and
\begin{align}\label{EW_couplings}
g_{L}(Z \rightarrow l^+l^-) & = -\frac{1}{2} + \sin^2 \theta_W \;, \nonumber \\
g_{R}(Z \rightarrow l^+l^-) & = \ \sin^2 \theta_W \;, \nonumber \\
g_{L}(Z \rightarrow q \bar{q}) & =T_3^q - e_q \sin^2 \theta_W \;,\nonumber \\
g_{R}(Z \rightarrow q \bar{q}) & =  - e_q \sin^2 \theta_W \;,
\end{align}
with $T_3^q$ being third component of the quark isospin, $e_q$ being
quark electric charge (in units of $|e|$).

The graviton contribution comes from both quark-antiquark
annihilation and gluon-gluon fusion subprocesses (see, for instance,
\cite{Giudice:05})
\begin{eqnarray} \label{parton_cross_sec}
\frac{d\sigma^{\mathrm{grav}}}{d\hat{t}}(q \bar{q} \rightarrow
l^+l^-) &=& \frac{\hat{s}^4 + 10\hat{s}^3 \hat{t} + 42 \, \hat{s}^2
\hat{t}^2 + 64 \hat{s} \, \hat{t}^3 + 32 \, \hat{t}^4}{1536 \, \pi
\hat{s}^2} \left| \mathcal{S}(\hat{s}) \right|^2 \;,
\nonumber \\
\frac{d\sigma^{\mathrm{grav}}}{d\hat{t}}(gg \rightarrow l^+l^-) &=&
-\frac{\hat{t}(\hat{s} + \hat{t}) (\hat{s}^2 + 2 \hat{s}\,\hat{t} +
2\,\hat{t}^2)}{256 \, \pi \hat{s}^2} \left|\mathcal{S}(\hat{s})
\right|^2
 \;,
\end{eqnarray}
where
\begin{equation}\label{KK_summation}
\mathcal{S}(s) = \frac{1}{\Lambda_{\pi}^2} \sum_{n=1}^{\infty}
\frac{1}{s - m_n^2 + i \, m_n \Gamma_n} \;
\end{equation}
is the invariant part of the partonic matrix elements, with
$\Gamma_n$ being \emph{total width of the graviton} with the KK
number $n$ and mass $m_n$~\cite{Kisselev:06}:
\begin{equation}\label{graviton_widths}
\Gamma_n = \eta \, m_n \left( \frac{m_n}{\Lambda_{\pi}} \right)^2,
\quad \eta \simeq 0.09 \;.
\end{equation}
Let us note that the function $\mathcal{S}(s)$ is the same for all
processes mediated by $s$-channel virtual gravitons.

In the RSSC model, an explicit form for the sum (\ref{KK_summation})
was obtained for $\bar{M}_5 \gg \kappa$ in Ref. \cite{Kisselev:06}
\begin{equation}\label{KK_sum}
\mathcal{S}(s) = - \frac{1}{4 \bar{M}_5^3 \sqrt{s}} \; \frac{\sin 2A
+ i \sinh 2\varepsilon }{\cos^2 \! A + \sinh^2 \! \varepsilon} \;,
\end{equation}
where
\begin{equation}\label{parameters}
A = \frac{\sqrt{s}}{\kappa} \;, \qquad \varepsilon = \frac{\eta}{2}
\Big( \frac{\sqrt{s}}{\bar{M}_5} \Big)^3 \;.
\end{equation}

It is important to underline that the magnitude of $\mathcal{S}(s)$
is defined by the fundamental gravity scale $\bar{M}_5$, not by the
scale $\Lambda_{\pi}$ presented in the Lagrangian. In general, this
property is valid in the RSSC model for \emph{both real and virtual}
production of the KK gravitons \cite{Kisselev:05,Kisselev:06}.




\end{document}